# EVALUATION OF THE FEYNMAN'S PROPAGATOR BY MEANS OF THE QUANTUM HAMILTON-JACOBI EQUATION

Mario Fusco Girard

*Department of Physics "E. R. Caianiello"*

*University of Salerno (Italy)*

*Electronic address: mfuscogirard@gmail.com*

ABSTRACT

It is shown that the complex phase of the Feynman's propagator is a solution of the quantum Hamilton-Jacobi equation, i.e. it is the quantum Hamilton's principal function (or quantum action). The propagator so can be computed either by means of the path integration, or by the way of that equation. This is analogous to what happens in classical mechanics, where the Hamilton's principal function can be computed or by integrating the lagrangian along the extremal paths, or as solution of a partial differential equation, namely the classical Hamilton-Jacobi equation. If the path is decomposed in the classical one and quantum fluctuations, the contribution of these latter satisfies a non-linear partial differential equation, whose coefficients depend on the classical action. When the contribution of the quantum fluctuations depend only on the time, it can be computed by means of a simple integration. The final results for the propagators in this case are equal to the Van Vleck-Pauli-Morette expressions, even though the two derivations are quite different.

## 1. THE METHOD

As well known, the propagator is the fundamental quantity in the Feynman's space-time approach to non-relativistic quantum mechanics [1, 2]. It gives the quantum amplitude for a particle to go from a point $x_A$ at time $t_A$, to $x_B$ at time $t_B$, as a path integral, i.e. a sum of contributions $\phi[x(t)]$ from each path connecting the two points in the time $t_B - t_A$ (for simplicity we adopt a one-dimensional notation):

$$K(x_B, t_B \mid x_A, t_A) = \int_{x_A, t_A}^{x_B, t_B} e^{\frac{i}{\hbar} \int_{t_A}^{t_B} L(x(t), v(t), t)\, dt}\, D[x(t)]$$

(1),

here $L(x(t), v(t), t)$ is the classical Lagrangian.

The propagator, originally named the kernel by Feynman, is the Green function of the Schrödinger's equation [2 - 6] :

$$\left( i\hbar \frac{\partial}{\partial t_B} - H(x_B, t_B) \right) K(x_B, t_B \mid x_A, t_A) = i\hbar\, \delta(x_B - x_A)\, \delta(t_B - t_A)$$

(2),

where H is the Hamiltonian operator.

When $x_B \neq x_A$ and $t_B > t_A$, the propagator therefore satisfies the time-dependent Schrödinger's equation, and in this way it can be connected to the Quantum Hamilton-Jacobi Equation (QHJE) [7, 8]. This latter, fully equivalent to the Schrödinger's equation, is the starting point of the WKB approximation, and appears when one tries to find solutions in exponential form

$$\psi(x, t) = C\, e^{\frac{i}{\hbar} S(x, t)}$$

(3)

of the Schrödinger's equation for a particle of mass m in a potential V(x):

$$i\hbar \frac{\partial \psi}{\partial t} = -\frac{\hbar^2}{2m} \frac{\partial^2 \psi}{\partial x^2} + V(x)\, \psi$$

(4).

$S(x, t)$ in (3) is a complex function and C is a constant.

By inserting (3) into (4), the time-dependent QHJE results:

$$\frac{\partial S(x,t)}{\partial t} + V(x) + \frac{1}{2m}\left(\frac{\partial S(x,t)}{\partial x}\right)^2 - \frac{i\hbar}{2m}\frac{\partial^2 S(x,t)}{\partial x^2} = 0 \qquad (5).$$

When $\hbar = 0$, last equation reduces to the classical Hamilton-Jacobi equation for the Hamilton's principal function S, also named the action [9].

As the propagator actually is a kind of wave function, this argument applies to it too. Therefore, if we write the propagator in the exponential form

$$K(x_B, t_B | x_A, t_A) = C\, e^{\frac{i}{\hbar} S(x_B, t_B | x_A, t_A)} \qquad (6),$$

where C is a constant, for $x_B \neq x_A$ and $t_B > t_A$, the quantum action $S(x_B, t_B | x_A, t_A)$ has to satisfy the eq. (5) with respect to $(x_B, t_B)$. Thus, if one finds the appropriate solution of (5), by means of (6) it is possible to compute the kernel $K(x_B, t_B | x_A, t_A)$ without do recourse to the path integration. Conversely, from the logarithm of a known propagator, the corresponding solution of (5) can be obtained.

A further step can be done by separating the path x(t) as

$$x(t) = x_{cl}(t) + y(t) \qquad (7),$$

where $x_{cl}(t)$ is the classical extremal path, and $y(t_B) = y(t_A) = 0$. The quantum action S is therefore split as

$$S(x_B, t_B | x_A, t_A) = S_{cl}(x_B, t_B | x_A, t_A) + \Delta(x_B, t_B | x_A, t_A) \qquad (8),$$

where $S_{cl}$ is the classical action

$$S_{cl}(x_B, t_B | x_A, t_A) = \int_{t_A}^{t_B} L(x_{cl}(t), v_{cl}(t), t)\, dt \qquad (9),$$

and the additive term $\Delta(x_B, t_B | x_A, t_A)$ gives the contributions from the expansion of $S(x_B, t_B | x_A, t_A)$ in terms of $y(t)$ and $\frac{dy}{dt}$ ( except for the linear one, which vanishes due to the motion's equation).

As a function of $(x_B, t_B)$, the classical action (9) satisfies the classical HJ equation (eq. 5 with $\hbar = 0$). Therefore, by inserting (8) with $x \equiv x_B$ and $t \equiv t_B$ considered as variables and fixed $x_A, t_A$, into (5), one gets an equation for the function $\Delta(x, t | x_A, t_A)$

$$\frac{\partial \Delta}{\partial t} + \frac{1}{2m}\left(2\frac{\partial S_{cl}}{\partial x}\frac{\partial \Delta}{\partial x} + \left(\frac{\partial \Delta}{\partial x}\right)^2\right) - \frac{i\hbar}{2m}\left(\frac{\partial^2 S_{cl}}{\partial x^2} + \frac{\partial^2 \Delta}{\partial x^2}\right) = 0$$

(10).

The known quantities in this equation depend on the classical action.

In the following, the dependence of the various functions on the parameters $(x_A, t_A)$ will be sometimes understood.

When $\Delta$ depends only on t, which happens for instance when the lagrangian is quadratic, the equation (10) reduces to

$$\frac{\partial \Delta}{\partial t} - \frac{i\hbar}{2m}\frac{\partial^2 S_{cl}}{\partial x^2} = 0$$

(11),

So that the function $\Delta(t)$ in this case is obtained as

$$\Delta(t) = \frac{i\hbar}{2m}\int_{t_A}^{t}\frac{\partial^2 S_{cl}}{\partial x^2}dt$$

(12).

Putting this into (8), the propagator is given by (6), apart from the multiplicative constant C which, following Feynman [2], can be computed by remembering that the kernel has to satisfy the equation

$$\psi(x_B, t_B) = \int_{-\infty}^{\infty} K(x_B, t_B | x_A, t_A)\psi(x_A, t_A)dx_A$$

(13).

The expansion of this equation in the quantities $\epsilon = t_B - t_A$ and $\eta = x_B - x_A$ gives

$$\psi(x, t) + \epsilon \frac{\partial \psi}{\partial t} = \int_{-\infty}^{\infty} K(x + \eta, t + \epsilon \mid x, t) \left( \psi(x, t) + \eta \frac{\partial \psi}{\partial x} + \eta^2 \frac{\partial^2 \psi}{\partial x^2} \right) d\eta$$

(14),

where the notation has been simplified by writing (x, t) instead of ($x_A$, $t_A$). By doing the integration and comparing the leading terms in $\epsilon$ on both sides, the constant C can be computed.

2. APPLICATIONS

a) *The free particle.*

The classical action is:

$$S_{cl}(x, t \mid x_A, t_A) = \frac{1}{2} \frac{m(x - x_A)^2}{(t - t_A)} \quad (15).$$

From (12) one gets:

$$\Delta(t) = \frac{i\hbar}{2} \text{Log}[t - t_A] \quad (16).$$

According to (8) the quantum action, i.e. the phase of the kernel, therefore is:

$$S(x, t \mid x_A, t_A) = \frac{1}{2} m \frac{(x - x_A)^{\wedge}2}{(t - t_A)} + i\hbar \, \text{Log}\left[\sqrt{t - t_A}\right] \quad (17),$$

and from (6), the kernel itself

$$K(x, t \mid x_A, t_A) = C \exp\left[\frac{i}{\hbar}(S_{cl}[x, t] + \Delta[t])\right] = C \frac{\exp\left[\frac{i}{\hbar} \frac{1}{2} m \frac{(x-x_A)^{\wedge}2}{(t - t_A)}\right]}{\sqrt{t - t_A}}$$

(18).

At the leading order in $\epsilon$, the eq. (14) gives

$$\psi(x, t) = C \frac{\int_{-\infty}^{\infty} e^{\frac{i m \eta^2}{2 \hbar \epsilon}} d\eta}{\sqrt{\epsilon}} \psi(x, t) \quad (19),$$

so that

$$C = \sqrt{\frac{m}{2 \pi i \hbar}} \quad (20),$$

which inserted into the last equality in (18), reproduces the correct result [2].

b) *The harmonic oscillator.*

The classical action is:

$$S_{cl}(x, t) = \frac{m \omega}{2 \sin[\omega(t - t_A)]} \left[ (x - x_A)^2 \cos[\omega(t - t_A)] - 2 x x_A \right] \quad (21).$$

From (12)

$$\Delta(t) = \frac{i \hbar}{2} \log[\sin[\omega(t - t_A)]] \quad (22).$$

The quantum action therefore is

$$S(x, t) = \frac{1}{2} m \omega x^2 \cot[\omega(t - t_A)] + \frac{1}{2} i \hbar \log[\sin[\omega(t - t_A)]] \quad (23),$$

and the kernel is evaluated as

$$K(x, t | x_A, t_A) = C \frac{\exp\left[\frac{i}{\hbar} \frac{m \omega}{2 \sin[\omega(t - t_A)]} \left[ (x - x_A)^2 \cos[\omega(t - t_A)] - 2 x x_A \right]\right]}{\sqrt{\sin[\omega(t - t_A)]}}$$

$$(24).$$

The constant C in this case is

$$C = \sqrt{\frac{m\omega}{2\pi i \hbar}} \quad (25).$$

With this value of the multiplicative constant, (24) gives the well-known propagator for the harmonic oscillator. The procedure above is simpler than the original one by Feynman [2], which integrates on the coefficients of the Fourier expansion for the kernel, and also with respect to that reported in many books [3-6], which exploits the limit of a difference equation.

c) *The driven harmonic oscillator.*

For the harmonic oscillator driven by a time-dependent force, the classical action is given by Eq. (3.66) of Ref. [2], and it is the sum of the corresponding one for the free harmonic oscillator, plus terms which are linear in the coordinates $x_A$ and $x_B$. Therefore, the function $\Delta(t)$ and the constant C are the same as for the undriven case, given by (22) and (25), respectively. In this case too, the propagator from our method is the same as computed by means of the path integration [4].

d) *Quadratic Lagrangian with time-depending coefficients.*

The method presented in Sect. 1 can be applied to the generalization of the previous case, i.e. when all the coefficients of the quadratic lagrangian depend on the time

$$L = \frac{1}{2}\left[a(t) v^2 - b(t) x^2\right] + c(t) x \quad (26).$$

In the following we will present the case of a damped oscillator [4], with the coefficients

$$a(t) = m e^{\gamma t}, \quad b(t) = m \omega^2 e^{\gamma t}, \quad c(t) = 0 \quad (27).$$

The classical lagrangian is

$$S_{cl}[x, t] = \frac{1}{2} \frac{m}{\sin[\Omega (t - t_A)]} \left( \frac{1}{2} \left( e^{\gamma t_A} x_A^2 - e^{\gamma t} x^2 \right) \gamma \right.$$

$$\left. + \Omega \left( \left( e^{\gamma t_A} x_A^2 + e^{\gamma t} x^2 \right) \cos[\Omega (t - t_A)] - 2 e^{\frac{1}{2} \gamma (t+t_A)} x x_A \right) \right)$$

(28),

where

$$\Omega = \sqrt{\omega^2 - \frac{\gamma^2}{4}}$$

(29).

The additive imaginary part $\Delta (t)$ to the quantum action is

$$\Delta (t) = -\frac{1}{4} i \hbar \gamma (t - t_A) + \frac{1}{2} i \hbar \log[\sin[\Omega (t_B - t_A)]]$$

(30).

Finally, by computing the multiplicative constant C by the method previously exposed, the propagator for this case results

$$K[x, t \mid x_A, t_A] = e^{\frac{t_A}{2} \gamma} \sqrt{\frac{m \Omega}{2 \pi i \hbar}} \frac{e^{\frac{(t-t_A) \gamma}{4}}}{\sqrt{\sin[\Omega (t - t_A)]}}$$

$$\exp \left[ \frac{i}{\hbar} \left( \frac{1}{2} m \Omega \left( \frac{1}{\sin[\Omega (t - t_A)]} \left( \left( e^{\gamma t_A} x_A^2 + e^{t \gamma} x^2 \right) \cos[\Omega (t - t_A)] - 2 e^{\frac{1}{2} \gamma (t+t_A)} x x_A \right) \right) + \right.\right.$$

$$\left.\left. \frac{1}{4} m \gamma \left( e^{\gamma t_A} x_A^2 - e^{\gamma t} x^2 \right) \right) \right]$$

(31),

which is the correct expression [4].

## 3. COMMENTS

In this paper we analyse the link between the Feynman's propagator and the quantum Hamilton-Jacobi equation. When the propagator is written in exponential form, its complex phase is a solution of the QHJE, with respect to the coordinates and time of the final point, and contains the coordinates and time of the initial point as parameters. By analogy with the classical case, we name it the quantum Hamilton principal function or quantum action. The propagator so can be computed either by means of the Feynman's path integration, or by means of that equation. This is analogous to what happens in classical mechanics, where the Hamilton's principal function can be computed or by integrating the lagrangian with respect to the time along each extremal path, in a region covered by a family of non-intersecting extremal paths [10], or as solution of a partial differential equation, i.e. the classical Hamilton-Jacobi equation.

In classical mechanics the action is a fundamental dynamical quantity. The same happens in the quantum case, being the corresponding quantity, i.e. the quantum action, the phase of the wave function, or the phase of the propagator, as the case. These quantum functions generate the corresponding classical ones in the limit when the Planck's constant h -> 0.

If the path is decomposed in the classical one and the quantum fluctuations, the contribution of these latter satisfies the non-linear partial differential equation (10), whose coefficients depend on the classical action. While this latter is a real function, the quantum one is a complex quantity. When the Planck's constant h goes to zero, the imaginary part of the quantum action vanishes and this function reduces to the classical corresponding one. As the Hamilton-Jacobi formulation of the classical mechanics is fully equivalent to those respectively based on the Lagrangian or the Hamiltonian functions, this shows how in this approach the classical mechanics emerges from the quantum one. In fact, these three formulations of the classical mechanics are all based on the same fundamental principle, i.e. the Hamilton's principle of least action.

When the contribution of the quantum fluctuations depend only of the time, as happens for the quadratic lagrangians, it is computed by means of a simple integration. It this case, the final result for the propagator is the same of the so called Van-Vleck-Pauli-Morette expression [11-13], even if the procedure is quite different.

REFERENCES

[1] R. P. Feynman, "*Space-Time Approach to Non-relativistic Quantum Mechanics*", Rev. Mod. Phys. 20 , 367 (1948).

[2] R. P. Feynman and A. R. Hibbs, "*Quantum Mechanics and Path Integrals*", McGraw-Hill, New York, 1965, and R. P. Feynman and A. R. Hibbs, "*Quantum Mechanics and Path Integrals*", Emended edition by D. F. Styer, McGraw-Hill, New York, 2005.

[3] L. S. Schulman, " *Techiques and Application of Path Integration*", John Wiley&Sons, New York, 1981.

[4] D. C. Khandekar, S. V. Lawande, K. V. Bhagwat, "*Path-Integral Methods and their Applications*", World Scientific, Singapore, 1993.

[5] C. Grosche, F. Steiner, " *Handbook of Feynman Path Integrals*", Springer-Verlag, Berlin, 1998.

[6] H. Kleinert, "*Path Integrals in Quantum Mechanics, Statistics, Polymer Physics, and Financial Markets",* 3th edition, World Scientific, Singapore, 2004.

[7] A. Messiah, "Quantum Mechanics", North Holland, Amsterdam (1961).

[8] M. Fusco Girard, "Analytical Solutions of the Quantum Hamilton-Jacobi Equation and Exact WKB-like Representations of One-Dimensional Wave Functions", arXiv: 1512.01356v1 [quant-ph] (2015).

[9] H. Goldstein, "Classical Mechanics", Reading Mass, Addison-Wesley 1950.

[10] M. Giaquinta and S. Hildebrandt, "*Calculus of Variations",* Voll. I and II, Springer Berlin Heidelberg 1996.

[11] J. H. Van Vleck, Proc. Nat. Acad. Sci. (USA) **14**, 178 (1928).

[12] W. Pauli, "*Selected Topics in Quantum Mechanics"* Cambridge, Mass, 1973.

[13] C. De Witt-Morette, Phys. Rev. **81**, 848 (1951).